\documentclass[conf]{new-aiaa}
\usepackage[utf8]{inputenc}
\usepackage{amsmath}
\usepackage{algorithmic}
\usepackage{graphicx}
\usepackage{textcomp}
\usepackage{xcolor}
\usepackage{multirow}
\usepackage{siunitx}
\usepackage{float}
\usepackage{subcaption}

\usepackage{tikz}
\usetikzlibrary{arrows.meta,calc,angles,quotes}
\tikzset{every picture/.style={line width=0.75pt}} %set default line width to 0.75pt  

\usepackage{cleveref}
\graphicspath{ {./images/}{./SCITECH/} }

\allowdisplaybreaks

\begin{document}

\title{Terrain-Aided Navigation Without Raycasting: A Sliding Grid Approach for Point Cloud Processing}

\author{Abdülbaki Şanlan\footnote{Research Assistant, Faculty of Aeronautics and Astronautics, sanlan19@itu.edu.tr (Corresponding Author)} and Fatih Erol\footnote{Research Assistant, Aerospace Research Center, erolfa@itu.edu.tr}}
\affil{Faculty of Aeronautics and Astronautics, Istanbul Technical University, Istanbul, Türkiye}
\author{Murad Abu-Khalaf}
\affil{Turkish Aerospace, Istanbul, Türkiye}
\author{Emre Koyuncu\footnote{Assoc. Prof., Faculty of Aeronautics and Astronautics, Aerospace Research Center, emre.koyuncu@itu.edu.tr}}
\affil{Faculty of Aeronautics and Astronautics, Istanbul Technical University, Istanbul, Türkiye}

\maketitle

\begin{abstract}
We investigate the use of a point cloud measurement in terrain-aided navigation. Our goal is to aid an inertial navigation system, by exploring ways to generate a useful measurement residual for effective nonlinear state estimation. We compare two such measurement models that involve the scanning of a digital terrain elevation model: a) one that is based on typical ray-casting from a given pose, that returns the predicted point cloud measurement from that pose, and b) another computationally less intensive one that does not require raycasting and we refer to herein as a sliding grid. Besides requiring a pose, it requires the pattern of the point cloud measurement itself and returns a predicted point cloud measurement. We show that the sliding grid achieves comparable accuracy to raycasting, especially at higher altitudes, while being orders of magnitude faster computationally. By eliminating the need for raycasting, the sliding grid removes the dependency on dedicated GPU hardware or high-performance CPUs, making it well-suited for real-time terrain-aided navigation on resource-constrained embedded platforms.
\end{abstract}


\section*{Nomenclature}

{\renewcommand\arraystretch{1.0}
\noindent\begin{tabular}{@{}l @{\quad=\quad} l}
$a$             & acceleration vector, $\mathbb{R}^3$ \\
$a_m$           & accelerometer measurement \\
$a_{\text{bias}}$ & accelerometer bias, $\mathbb{R}^3$ \\
$c$             & received point cloud measurement \\
$\hat{c}$       & predicted point cloud measurement (raycasting) \\
$\hat{c}_z$     & predicted point cloud $z_s$-component (sliding grid) \\
$c_{[xyz]}$     & $x$--$y$--$z$ pattern of received point cloud \\
$d_a,\, d_{\omega}$ & Gaussian bias drift inputs \\
$g$             & gravity vector, $\mathbb{R}^3$ \\
$M$             & number of point cloud measurement points \\
$N$             & number of particles \\
$n_a$           & accelerometer measurement noise \\
$n_s$           & point cloud measurement noise \\
$n_{\omega}$    & gyroscope measurement noise \\
$p$             & position vector, $\mathbb{R}^3$ \\
$q$             & orientation quaternion \\
$r$             & measurement residual \\
$R_b^i$         & rotation matrix from body-frame to inertial-frame \\
$v$             & velocity vector, $\mathbb{R}^3$ \\
$w_m$           & gyroscope measurement \\
\end{tabular}}

\vspace{4pt}
{\renewcommand\arraystretch{1.0}
\noindent\begin{tabular}{@{}l @{\quad=\quad} l}
$\delta p$      & position error-state \\
$\delta v$      & velocity error-state \\
$\delta \theta$  & attitude error-state \\
$\mu$           & measurement model function \\
$\omega$        & angular velocity, $\mathbb{R}^3$ \\
$\omega_{\text{bias}}$ & gyroscope bias, $\mathbb{R}^3$ \\
$\otimes$       & Hamilton quaternion product \\
\end{tabular}}

\vspace{4pt}
{\renewcommand\arraystretch{1.0}
\noindent\begin{tabular}{@{}l @{\quad=\quad} l}
$\mathcal{D}$   & digital elevation model (DEM) \\
$\mathcal{F}_b$ & body-frame \\
$\mathcal{F}_i$ & inertial-frame (local navigation frame) \\
$\mathcal{S}$   & sensor scan parameters \\
$\mathcal{T}$   & terrain surface \\
$[\,\cdot\,]_{\times}$ & skew-symmetric (cross-product) matrix \\
\end{tabular}}

\vspace{4pt}
{\renewcommand\arraystretch{1.0}
\noindent\begin{tabular}{@{}l @{\quad=\quad} l}
$\overline{(\cdot)}$   & nominal-state quantity \\
$\delta(\cdot)$        & error-state quantity \\
$\hat{(\cdot)}$        & estimated or predicted quantity \\
\end{tabular}}

\section{Introduction}
Terrain-aided navigation has been mostly based on altimeter measurements, where significant progress has been made since TERCOM \cite{Golden_TERCOM_1980} in the 1950s and SITAN \cite{SITAN_HostetlerAndreas_1983} in the 1970s. In TERCOM \cite{Golden_TERCOM_1980}, a database of separate patches of above mean sea level (MSL) terrain elevation data are placed along the planned track of an aircraft and are loaded on board. This on-board data is matched and correlated with a sequence of above MSL terrain elevation measurements of the executed flight track, using a radar altimeter and a barometric altimeter. This generates a position fix that is used by a state estimator to improve the navigation solution, which is then used for guidance to plan any necessary course adjustments to be executed. SITAN \cite{SITAN_HostetlerAndreas_1983} on the other hand uses, attempts to utilize fully nonlinear Kalman filtering techniques by using several local linearization mechanisms of the measurement model to use by an extended Kalman filter. With the advent of particle filtering in the early 1990s \cite{Gordon_ParticleFilter_1993}, the need to linearize a highly nonlinear measurement model is circumvented as seen in results such as \cite{SchonMarginalized2005,NordlundMarginalized2009} who used marginalized particle filters and altimeter measurements. In \cite{Turan_ParticleFilterTAN,Turan_PFUnscentedTAN,TuranFlightTest2023}, particle filters were to process altimeter information together with a flight test of the used methodology.

Beyond the use of an altimeter, two airborne laser scanners were tested in \cite{VadlamaniDualAirborneLaser2009} to generate two digital elevation maps whose displacement is used to aid the INS in a dead-reckoning manner. In \cite{UijtdeHaag_Laserrange_2006}, point cloud measurements made up of a large number of terrain samples is compared against scanned Digital Elevation Model (DEM) \cite{national_geospatial-intelligence_agency_digital_2022} data by minimizing a Sum of Squared Error (SSE) metric. The scanned DEM data is given an attitude estimation for a matrix of offset positions from the current best location estimate; alternatively the authors proposed minimizing the error surface of SSE via gradient-based search methods. In \cite{Steereable-laser-MPF}, a steerable-laser measurement sensor that is optimally directed at terrain information that can maximize an information matrix, and uses marginalized particle filters. The author also compares the steerable-laser against a three laser tri-pod configuration. Alongside these range-based methods, camera-based GNSS-denied navigation has been investigated using satellite image feature matching \cite{csanlan2025particle} and triangulation aware visual inertial odometry \cite{csanlan2026tango}.

The objective of this study is to investigate the use of a point cloud measurement for terrain-aided navigation, where an aircraft is equipped with a strapdown inertial navigation system (INS), a measurement sensor that generates a point cloud of the surrounding terrain, and a digital elevation model (DEM) database. The point cloud measurement size ($M$) exceeds the single-point returned by a steerable-laser sensor or the 3-point tripod configuration, yet it does not need to be as large as the dense airborne laser scanner measurement used in \cite{UijtdeHaag_Laserrange_2006}. Since both the strapdown motion prediction model and the point cloud measurement are highly nonlinear functions of the states and the terrain, respectively, the central objective is to develop an efficient measurement model that generates a useful measurement residual from the difference between the received point cloud of the real terrain and a corresponding scan of the available DEM, while keeping the computational burden compatible with resource-constrained embedded platforms.

Reformulating the measurement model offers one way to meet this computational objective, as demonstrated in range-based localization \cite{thrun2001probabilistic,biber2003normal} and visual--inertial navigation \cite{mourikis2007multi}. Following this principle, we propose a novel measurement model for terrain-aided navigation that we refer to as a \emph{sliding grid}. Unlike conventional raycasting, in which predicted measurements are generated by iteratively casting rays from a hypothetical aircraft pose and finding surface intersections on the DEM---a process that is computationally intensive and typically requires GPU acceleration or significant CPU resources---the sliding grid leverages the spatial pattern of the already received point cloud measurement. Specifically, for each hypothetical aircraft pose represented by a particle in the filter, the method reuses the sensor-frame $x_s$--$y_s$ pattern of the received point cloud to compute the corresponding inertial-frame horizontal coordinates, and then performs a direct orthogonal lookup of the DEM elevation at those coordinates to obtain the predicted $z_s$ values. This transforms the computationally expensive three-dimensional ray--surface intersection problem into a simple two-dimensional table lookup. In sliding grid mode, therefore, the $x_s$--$y_s$ pattern is fixed across all particles and matching relies on the $z_s$ values only, whereas raycasting offers the flexibility to match across all three dimensions ($x_s$, $y_s$, $z_s$) or subsets thereof. For nonlinear state estimation, we employ a marginalized particle filter with $N$ particles to estimate the error states of the INS and correct its nominal state in an open-loop architecture. Each particle carries an associated scan of the digital terrain that is compared with the received point cloud to compute the likelihood, resulting in $N \times M$ scan points that must be evaluated at every measurement update. The sliding grid is particularly advantageous in this context, as it eliminates the raycasting bottleneck for all $N \times M$ evaluations. The sliding grid method is especially well-suited for platforms with limited onboard computing power, such as small unmanned aerial vehicles, or for operational regimes where rapid processing is prioritized. 

The contributions of this work are as follows. First, whereas the point cloud--based terrain matching in \cite{UijtdeHaag_Laserrange_2006} relies on a dense airborne laser scanner measurement and minimizes a Sum of Squared Error metric via gradient-based search over a grid of offset positions, this study employs a moderately sized point cloud within a marginalized particle filter framework, demonstrating that effective terrain-aided navigation does not require an excessively large number of terrain samples. Second, while prior marginalized particle filter formulations have been limited to scalar altimeter measurements \cite{SchonMarginalized2005,NordlundMarginalized2009} or single-point steerable-laser returns \cite{Steereable-laser-MPF}, and all existing point cloud--based approaches rely on computationally expensive raycasting that demands dedicated GPU hardware or high-performance CPUs \cite{UijtdeHaag_Laserrange_2006,Steereable-laser-MPF,VadlamaniDualAirborneLaser2009}, the present work extends the marginalized particle filter to process a multi-point cloud measurement through the proposed sliding grid---a novel raycasting-free measurement model that achieves comparable localization accuracy, especially at cruise altitudes, while being orders of magnitude faster computationally, thereby enabling real-time terrain-aided navigation on resource-constrained embedded platforms without specialized hardware.

The paper is organized as follows. In \Cref{Section:Preliminaries}, we discuss the utilized sensors and the decomposition of the total-state navigation kinematics into a nominal-state and an error-state kinematics. In \Cref{Section:MeasurementModel}, we present the two terrain measurement models considered: raycasting and the proposed sliding grid. In \Cref{Section:StateEstimator}, we describe the marginalized particle filter used for state estimation. In \Cref{Section:Results}, we report simulation results comparing the two measurement models across several scenarios. In \Cref{Section:Conclusion} we provide conclusions.

\section{Preliminaries}\label{Section:Preliminaries}
This section establishes the sensor suite and the motion prediction model that underpin the state estimator developed in this work. We first describe the on-board sensors assumed to be available, and then present the total-state inertial navigation kinematics along with its decomposition into nominal-state and error-state representations, which forms the basis for the marginalized particle filter formulation introduced in \Cref{Section:StateEstimator}.

\subsection{Utilized Sensors}

We assume the aircraft is equipped with a navigation-grade Inertial Measurement Unit (IMU) \cite[Chapter~4]{grove2013principles}, which provides three-axis accelerometer and gyroscope measurements at a high update rate, and a terrain-scanning sensor capable of generating a point cloud measurement of the surrounding terrain surface. The point cloud sensor may be realized, for instance, by a LiDAR or a multi-beam laser scanner. No radar altimeter or barometric altimeter is used; the only external aiding source is the point cloud measurement together with an on-board digital elevation model database.

\subsection{Motion Model}\label{Section:MotionModel}
Inertial navigation mechanization equations that generate a navigation solution require the use of a motion prediction model \cite{grove2013principles}. The Pinson model is one such model \cite{pinson1963inertial}\cite[Chapter 14]{grove2013principles} that models the dynamics of a drifting Inertial Navigation System (INS).

In this regard, consider the INS dynamics that reflects the total-state navigation kinematics \cite{sola2017quaternion}
\begin{subequations}\label{eq:INSDynamics_TotalStateKinematics}
\begin{align}
    \dot{p} &=v,\\
    \dot{v} &= a \nonumber\\
            &= R_{b}^{i} (a_m - a_{\text{bias}} - n_a) + g,\\
    \dot{q} &= \tfrac{1}{2} q \otimes w \nonumber\\
            &= \tfrac{1}{2} q \otimes (w_m -  \omega_{\text{bias}} - n_{\omega}),\\
    \dot{a}_{\text{bias}}&= d_a,\\
    \dot{\omega}_{\text{bias}}&= d_{\omega},\\
    \dot{g} &= 0,
\end{align}
\end{subequations}
where $p \in \mathbb{R}^3$, $v\in \mathbb{R}^3$ and $a\in \mathbb{R}^3$ are the true position, velocity and acceleration vectors of the aircraft resolved with respect to the axes of an inertial-frame $\mathcal{F}_i$, which for the purposes of this current study is assumed to be the local navigation frame. Moreover, $q$ is a quaternion that describes the orientation of the body-frame $\mathcal{F}_b$ with respect to $\mathcal{F}_i$ and $\otimes$ is the Hamilton product. The angular velocity $\omega \in \mathbb{R}^3$ is resolved with respect to the axes of $\mathcal{F}_b$. The 3-axis accelerometer bias is given by $a_{\text{bias}}\in \mathbb{R}^3$ whereas $\omega_{\text{bias}}\in \mathbb{R}^3$ is the gyroscope bias. The gravity vector $g\in \mathbb{R}^3$ is assumed to be constant in the region of interest. Lastly, it is assumed that the drift in the IMU bias characteristics is driven by Gaussian distribution inputs $d_a$ and $d_{\omega}$.

The output of the IMU 
\begin{subequations}\label{eq:IMUOutput}
\begin{align}
    a_m &= R_{i}^{b} (a - g ) + a_{\text{bias}} + n_a,\\
    w_m &= \omega + \omega_{\text{bias}} + n_{\omega}.
\end{align}
\end{subequations}
where $a_m$ is the acceleration measurement and $n_a$ is the associated measurement noise; $\omega_m$ is the angular rate measurement and $n_{\omega}$ is the associated measurement noise. $R_{i}^{b}$ is a rotation matrix and transforms the coordinates from $\mathcal{F}_i$ to $\mathcal{F}_b$.

To perform filtering and state estimation using the motion prediction model, it is helpful to decompose the total-state true navigation kinematics \eqref{eq:INSDynamics_TotalStateKinematics} into a nominal-state kinematics \eqref{eq:INSDynamics_NominalStateKinematics} and error-state kinematics \eqref{eq:INSDynamics_ErrorStateKinematics}. A state-estimator filter that uses terrain information can then be designed to obtain optimal estimates of the error-state vector. Estimating the error-state directly, rather than the total-state, is a way to overcome the nonlinearities inherent in the full navigation kinematics, and enables the design of computationally efficient marginalized particle filters as discussed in \Cref{Section:StateEstimator}.

The nominal-state results from the integration of the IMU readings, and thus represents the dead-reckoning navigation solution for the purposes of this current study and is given by \cite{sola2017quaternion}:
\begin{subequations}\label{eq:INSDynamics_NominalStateKinematics}
\begin{align}
    \dot{\bar{p}} &=\bar{v},\\
    \dot{\bar{v}} &= \bar{R}_{b}^{i} (a_m - \bar{a}_{\text{bias}}) + \bar{g},\\
    \dot{\bar{q}} &= \tfrac{1}{2} \bar{q} \otimes (w_m -  \bar{\omega}_{\text{bias}}),\\
    \dot{\bar{a}}_{\text{bias}}&= 0,\\
    \dot{\bar{\omega}}_{\text{bias}}&= 0,\\
    \dot{\bar{g}} &= 0.
\end{align}
\end{subequations}

Meanwhile, the error-state kinematics are given by \cite{sola2017quaternion}:
\begin{subequations}\label{eq:INSDynamics_ErrorStateKinematics}
\begin{align}
    \dot{\delta{p}} &=\delta{v},\\
    \dot{\delta{v}} &= \bar{R}_{b}^{i} \left(-[a_m - \bar{a}_{\text{bias}}]_{\times} \delta{\theta} -  \delta{a}_{\text{bias}} - n_a \right)+ \delta{g},\\
    \dot{\delta{\theta}} &= -[\omega_m - \bar{\omega}_{\text{bias}}]_{\times} \delta{\theta} - \delta{\omega}_{\text{bias}} -n_{\omega},\\
    \dot{\delta{a}}_{\text{bias}}&= d_a,\\
    \dot{\delta{\omega}}_{\text{bias}}&= d_{\omega},\\
    \dot{\delta{g}} &= 0.
\end{align}
\end{subequations}

In this large-signal/small-signal decomposition, a non-rotational quantity such as position is given by $p=\bar{p} +\delta p$, while $R = \bar{R}\delta R$ where $\delta R = \exp([\delta\theta]_{\times})$. Note that $[a]_{\times}b$ denotes the skew-symmetric matrix representation for cross product $a\times b$.

\section{Terrain Measurement Model}\label{Section:MeasurementModel}

The state estimator requires a terrain measurement model to filter and correct the predicted error-state using the likelihood of the received terrain measurement. The measurement prediction model is used to provide a measurement residual which compares the \emph{received} measurement against the \emph{predicted} measurement.

\begin{figure}[H]
    \centering
    \begin{tikzpicture}[x=0.75pt,y=0.75pt,yscale=-0.5,xscale=0.5]
    %uncomment if require: \path (0,542); %set diagram left start at 0, and has height of 542
    %Snip Round Single Corner Rect [id:dp028389949814091264] 
    \draw  [color={rgb, 255:red, 184; green, 233; blue, 134 }  ,draw opacity=1 ][fill={rgb, 255:red, 184; green, 233; blue, 134 }  ,fill opacity=1 ][dash pattern={on 3.75pt off 3pt on 7.5pt off 1.5pt}] (-1,369.5) .. controls (-1,356.52) and (9.52,346) .. (22.5,346) -- (175.5,346) -- (199,369.5) -- (199,463.5) -- (-1,463.5) -- cycle ;
    %Snip Round Single Corner Rect [id:dp3707273890517011] 
    \draw  [color={rgb, 255:red, 184; green, 233; blue, 134 }  ,draw opacity=1 ][fill={rgb, 255:red, 184; green, 233; blue, 134 }  ,fill opacity=1 ] (199,431.5) .. controls (199,427.08) and (202.58,423.5) .. (207,423.5) -- (691,423.5) -- (699,431.5) -- (699,463.5) -- (199,463.5) -- cycle ;
    %Rounded Rect [id:dp2842100066414748] 
    \draw  [fill={rgb, 255:red, 0; green, 0; blue, 0 }  ,fill opacity=1 ] (270.08,186.87) .. controls (270.55,185.93) and (271.7,185.55) .. (272.63,186.03) -- (365.65,232.99) .. controls (366.59,233.47) and (366.96,234.61) .. (366.49,235.54) -- (363.92,240.63) .. controls (363.45,241.57) and (362.3,241.95) .. (361.37,241.47) -- (268.35,194.51) .. controls (267.41,194.03) and (267.04,192.89) .. (267.51,191.96) -- cycle ;
    %Shape: Circle [id:dp9449770627301675] 
    \draw  [fill={rgb, 255:red, 74; green, 74; blue, 74 }  ,fill opacity=1 ] (304.5,213.75) .. controls (304.5,206.85) and (310.1,201.25) .. (317,201.25) .. controls (323.9,201.25) and (329.5,206.85) .. (329.5,213.75) .. controls (329.5,220.65) and (323.9,226.25) .. (317,226.25) .. controls (310.1,226.25) and (304.5,220.65) .. (304.5,213.75) -- cycle ;
    %Straight Lines [id:da31781872567722824] 
    \draw [color={rgb, 255:red, 208; green, 2; blue, 27 }  ,draw opacity=1 ]   (40,347.5) -- (317,213.75) ;
    %Straight Lines [id:da8640694294954594] 
    \draw [color={rgb, 255:red, 208; green, 2; blue, 27 }  ,draw opacity=1 ]   (317,213.75) -- (374,425.5) ;
    %Straight Lines [id:da00729587869971271] 
    \draw  [dash pattern={on 0.84pt off 2.51pt}]  (149,131.5) -- (455,282.5) ;
    %Straight Lines [id:da9084609059418862] 
    \draw [color={rgb, 255:red, 0; green, 0; blue, 255 }  ,draw opacity=1 ][line width=1.5]  [dash pattern={on 1.69pt off 2.76pt}]  (149,131.5) -- (40,347.5) ;
    %Straight Lines [id:da14221642925422673] 
    \draw [color={rgb, 255:red, 0; green, 0; blue, 255 }  ,draw opacity=1 ][line width=1.5]  [dash pattern={on 1.69pt off 2.76pt}]  (455,282.5) -- (374,425.5) ;
    
    % Text Node
    \draw (60,201.4) node [anchor=north west][inner sep=0.75pt]  [color={rgb, 255:red, 0; green, 0; blue, 255 }  ,opacity=1 ]  { $z_{1}{}_{s}$};
    % Text Node
    \draw (421,342.4) node [anchor=north west][inner sep=0.75pt]  [color={rgb, 255:red, 0; green, 0; blue, 255 }  ,opacity=1 ]  {$z_{2}{}_{s}$};
    % Text Node
    \draw (201,130.4) node [anchor=north west][inner sep=0.75pt]    {$y_{1}{}_{s}$};
    % Text Node
    \draw (400,230.4) node [anchor=north west][inner sep=0.75pt]    {$y_{2}{}_{s}$};
    % Text Node
    \draw (314.95,163.37) node [anchor=north west][inner sep=0.75pt]  [rotate=-359.63] [align=left] {aircraft true pose};
    % Text Node
    \draw (181.95,434.37) node [anchor=north west][inner sep=0.75pt]  [rotate=-359.63] [align=left] {terrain};
    \end{tikzpicture}
     \caption{Point cloud measurement in ZY plane observed by an aircraft scanning the terrain below.}
    \label{Figure:AircraftTruePose}
\end{figure}

This measurement, as seen in \Cref{Figure:AircraftTruePose}, represents distance(s) to the terrain surface from the current aircraft pose. It is a function of the aircraft position $p$, orientation quaternion $q$, the terrain surface $\mathcal{T}$, the sensor scan parameters $\mathcal{S}$ (e.g., beam directions, scan pattern, and field of view), and the measurement noise $n_s$. For a point cloud measurement, this is given by
\begin{equation}\label{Eq:pointCloudModel}
    c = \mu\left(p,\, q,\, \mathcal{T},\, \mathcal{S},\, n_s\right).
\end{equation}

To predict a measurement, given how arbitrary a terrain surface $\mathcal{T}$ is, there is no closed-form representation of the measurement model. Instead, the terrain is represented by a digital elevation model (DEM) $\mathcal{D}$ \cite{national_geospatial-intelligence_agency_digital_2022}, and a measurement is predicted computationally for a given hypothetical pose---composed of the nominal position $\bar{p}$, nominal orientation $\bar{q}$, estimated position error $\hat{\delta p}$, and estimated attitude error $\hat{\delta\theta}$---together with $\mathcal{D}$ and the scan parameters $\mathcal{S}$:
\begin{equation}\label{Eq:pointCloudPredictionModel}
    \hat{c} = \mu\left(\bar{p},\, \bar{q},\, \hat{\delta p},\, \hat{\delta\theta},\, \mathcal{D},\, \mathcal{S}\right).
\end{equation}

To get $\hat{c}$ in \cref{Eq:pointCloudPredictionModel} requires the use of raycasting algorithms, which GPUs can handle very efficiently. We have implemented two types of raycasting algorithms, one that is based on a ray-triangle intersection \cite{doi:10.1080/10867651.1997.10487468} which guarantees empty space between the ray source and the surface it hits. We also implemented one that is based on binary search or bi-sectioning, as suggested in \cite{Steereable-laser-MPF} and can be handled by CPUs, and whose speed and accuracy  depend on the number of bits or bisection steps, and a stopping criteria that ensures stopping at the first facing surface.

Moreover, to leverage the pattern of the received point cloud $c$, we implemented a novel measurement innovation that we propose herein which given a hypothetical pose of the aircraft, it predicts the $z$ coordinates in sensor-frame of the expected point cloud measurement without raycasting. This measurement model leverages the already existing $x_s-y_s$ pattern of the received point cloud measurement $c$ of \eqref{Eq:pointCloudModel} from the true pose of the aircraft as seen for a simplified example in \Cref{Figure:AircraftTruePose}. For any hypothetical aircraft location and using the $x_s-y_s$ pattern in hand, it then traverses along the $z_s$-axis for each point towards the received $z_{s1}$ reading. For this $(x_{1s},y_{1s},z_{1s})$ point, its inertial frame horizontal coordinates are calculated and then the DEM is read at that horizontal location to give us $\hat{z}_{1s}$. This is essentially bootstrapping from the known pattern by orthogonally looking up the closest terrain point of the DEM to generate $\hat{z}_s$ as seen in \Cref{Figure:AircraftHypotheticalPoseAbove}. It works whether the $z_{s1}$ reading happens to be above or below the DEM surface as seen in \Cref{Figure:AircraftHypotheticalPoseAbove,Figure:AircraftHypotheticalPoseBelow}.

In this case, the predicted $z_s$-component of the point cloud is given by
\begin{equation}\label{Eq:pointCloudZPredictionModel}
    \hat{c}_z = \mu\left(\bar{p},\, \bar{q},\, \hat{\delta p},\, \hat{\delta\theta},\, c_{[xyz]},\, \mathcal{D},\, \mathcal{S}\right).
\end{equation}

Here $c_{[xyz]}$ is written to emphasize the particular use of the $x-y-z$ pattern of the point cloud measurement $c$, and in an order-preserving manner. It can be seen that \cref{Eq:pointCloudZPredictionModel} will generate the correct $z$ values within measurement noise range if the hypothetical location is placed at the true location, and thus attaining minimum error of almost zero, as demonstrated in \Cref{fig:ErrorSurfaceComparison}.
   
\begin{figure}[!htb]
    \centering
    \begin{tikzpicture}[x=0.75pt,y=0.75pt,yscale=-0.5,xscale=0.5]
        %uncomment if require: \path (0,542); %set diagram left start at 0, and has height of 542
        
        %Snip Round Single Corner Rect [id:dp585947169289084] 
        \draw  [color={rgb, 255:red, 184; green, 233; blue, 134 }  ,draw opacity=1 ][fill={rgb, 255:red, 184; green, 233; blue, 134 }  ,fill opacity=1 ][dash pattern={on 3.75pt off 3pt on 7.5pt off 1.5pt}] (-1,369.5) .. controls (-1,356.52) and (9.52,346) .. (22.5,346) -- (175.5,346) -- (199,369.5) -- (199,463.5) -- (-1,463.5) -- cycle ;
        %Snip Round Single Corner Rect [id:dp34541454546409334] 
        \draw  [color={rgb, 255:red, 184; green, 233; blue, 134 }  ,draw opacity=1 ][fill={rgb, 255:red, 184; green, 233; blue, 134 }  ,fill opacity=1 ] (199,431.5) .. controls (199,427.08) and (202.58,423.5) .. (207,423.5) -- (691,423.5) -- (699,431.5) -- (699,463.5) -- (199,463.5) -- cycle ;
        %Rounded Rect [id:dp9467876383211957] 
        \draw  [fill={rgb, 255:red, 0; green, 0; blue, 0 }  ,fill opacity=1 ] (342.08,116.87) .. controls (342.55,115.93) and (343.7,115.55) .. (344.63,116.03) -- (437.65,162.99) .. controls (438.59,163.47) and (438.96,164.61) .. (438.49,165.54) -- (435.92,170.63) .. controls (435.45,171.57) and (434.3,171.95) .. (433.37,171.47) -- (340.35,124.51) .. controls (339.41,124.03) and (339.04,122.89) .. (339.51,121.96) -- cycle ;
        %Shape: Circle [id:dp9388558334594577] 
        \draw  [fill={rgb, 255:red, 74; green, 74; blue, 74 }  ,fill opacity=1 ] (376.5,143.75) .. controls (376.5,136.85) and (382.1,131.25) .. (389,131.25) .. controls (395.9,131.25) and (401.5,136.85) .. (401.5,143.75) .. controls (401.5,150.65) and (395.9,156.25) .. (389,156.25) .. controls (382.1,156.25) and (376.5,150.65) .. (376.5,143.75) -- cycle ;
        %Straight Lines [id:da27350975457697146] 
        \draw [color={rgb, 255:red, 208; green, 2; blue, 27 }  ,draw opacity=1 ]   (112,277.5) -- (389,143.75) ;
        %Straight Lines [id:da7645940063700974] 
        \draw [color={rgb, 255:red, 208; green, 2; blue, 27 }  ,draw opacity=1 ]   (389,143.75) -- (446,355.5) ;
        %Straight Lines [id:da37180587168309065] 
        \draw  [dash pattern={on 0.84pt off 2.51pt}]  (221,61.5) -- (527,212.5) ;
        %Straight Lines [id:da6903693419068034] 
        \draw [color={rgb, 255:red, 0; green, 0; blue, 255 }  ,draw opacity=1 ][line width=1.5]  [dash pattern={on 1.69pt off 2.76pt}]  (221,61.5) -- (143.03,216.02) -- (112,277.5) ;
        %Straight Lines [id:da5839668674293202] 
        \draw [color={rgb, 255:red, 0; green, 0; blue, 255 }  ,draw opacity=1 ][line width=1.5]  [dash pattern={on 1.69pt off 2.76pt}]  (527,212.5) -- (446,355.5) ;
        %Straight Lines [id:da06569354164137131] 
        \draw    (112,346.5) .. controls (110.33,344.83) and (110.33,343.17) .. (112,341.5) .. controls (113.67,339.83) and (113.67,338.17) .. (112,336.5) .. controls (110.33,334.83) and (110.33,333.17) .. (112,331.5) .. controls (113.67,329.83) and (113.67,328.17) .. (112,326.5) .. controls (110.33,324.83) and (110.33,323.17) .. (112,321.5) .. controls (113.67,319.83) and (113.67,318.17) .. (112,316.5) .. controls (110.33,314.83) and (110.33,313.17) .. (112,311.5) .. controls (113.67,309.83) and (113.67,308.17) .. (112,306.5) .. controls (110.33,304.83) and (110.33,303.17) .. (112,301.5) .. controls (113.67,299.83) and (113.67,298.17) .. (112,296.5) .. controls (110.33,294.83) and (110.33,293.17) .. (112,291.5) .. controls (113.67,289.83) and (113.67,288.17) .. (112,286.5) .. controls (110.33,284.83) and (110.33,283.17) .. (112,281.5) -- (112,277.5) -- (112,277.5) ;
        %Straight Lines [id:da44850373287588696] 
        \draw    (446,424.5) .. controls (444.33,422.83) and (444.33,421.17) .. (446,419.5) .. controls (447.67,417.83) and (447.67,416.17) .. (446,414.5) .. controls (444.33,412.83) and (444.33,411.17) .. (446,409.5) .. controls (447.67,407.83) and (447.67,406.17) .. (446,404.5) .. controls (444.33,402.83) and (444.33,401.17) .. (446,399.5) .. controls (447.67,397.83) and (447.67,396.17) .. (446,394.5) .. controls (444.33,392.83) and (444.33,391.17) .. (446,389.5) .. controls (447.67,387.83) and (447.67,386.17) .. (446,384.5) .. controls (444.33,382.83) and (444.33,381.17) .. (446,379.5) .. controls (447.67,377.83) and (447.67,376.17) .. (446,374.5) .. controls (444.33,372.83) and (444.33,371.17) .. (446,369.5) .. controls (447.67,367.83) and (447.67,366.17) .. (446,364.5) .. controls (444.33,362.83) and (444.33,361.17) .. (446,359.5) -- (446,355.5) -- (446,355.5) ;
        %Straight Lines [id:da09944684126203496] 
        \draw [color={rgb, 255:red, 189; green, 16; blue, 224 }  ,draw opacity=1 ][line width=1.5]  [dash pattern={on 1.69pt off 2.76pt}]  (250,77.5) -- (112,346.5) ;
        %Straight Lines [id:da6927729322466278] 
        \draw [color={rgb, 255:red, 189; green, 16; blue, 224 }  ,draw opacity=1 ][line width=1.5]  [dash pattern={on 1.69pt off 2.76pt}]  (555,227.5) -- (446,424.5) ;
        %Straight Lines [id:da3760335740636358] 
        \draw  [dash pattern={on 0.75pt off 0.75pt}]  (527.71,211.18) .. controls (529.96,210.49) and (531.43,211.28) .. (532.12,213.54) .. controls (532.8,215.79) and (534.27,216.58) .. (536.52,215.9) .. controls (538.77,215.22) and (540.24,216.01) .. (540.93,218.26) .. controls (541.62,220.51) and (543.09,221.3) .. (545.34,220.62) .. controls (547.59,219.94) and (549.06,220.73) .. (549.75,222.98) .. controls (550.43,225.23) and (551.9,226.02) .. (554.15,225.34) -- (555.71,226.18) -- (555.71,226.18)(526.29,213.82) .. controls (528.54,213.14) and (530.01,213.93) .. (530.7,216.18) .. controls (531.39,218.43) and (532.86,219.22) .. (535.11,218.54) .. controls (537.37,217.86) and (538.84,218.65) .. (539.51,220.91) .. controls (540.2,223.16) and (541.67,223.95) .. (543.92,223.27) .. controls (546.17,222.59) and (547.64,223.38) .. (548.33,225.63) .. controls (549.02,227.88) and (550.49,228.67) .. (552.74,227.99) -- (554.29,228.82) -- (554.29,228.82) ;
        
        % Text Node
        \draw (127,147.4) node [anchor=north west][inner sep=0.75pt]  [color={rgb, 255:red, 0; green, 0; blue, 255 }  ,opacity=0.93 ]  {$z_{1}{}_{s}$};
        % Text Node
        \draw (460,262.4) node [anchor=north west][inner sep=0.75pt]  [color={rgb, 255:red, 0; green, 0; blue, 255 }  ,opacity=1 ]  {$z_{2}{}_{s}$};
        % Text Node
        \draw (273,60.4) node [anchor=north west][inner sep=0.75pt]    {$y_{1}{}_{s}$};
        % Text Node
        \draw (472,160.4) node [anchor=north west][inner sep=0.75pt]    {$y_{2}{}_{s}$};
        % Text Node
        \draw (411,107) node [anchor=north west][inner sep=0.75pt]   [align=left] {aircraft hypothetical pose};
        % Text Node
        \draw (208,164.4) node [anchor=north west][inner sep=0.75pt]  [color={rgb, 255:red, 189; green, 16; blue, 224 }  ,opacity=1 ]  {$\hat{z}_{1}{}_{s}$};
        % Text Node
        \draw (502.5,329.4) node [anchor=north west][inner sep=0.75pt]  [color={rgb, 255:red, 189; green, 16; blue, 224 }  ,opacity=1 ]  {$\hat{z}_{2}{}_{s}$};
        % Text Node
        \draw (184,437) node [anchor=north west][inner sep=0.75pt]   [align=left] {DEM};
    \end{tikzpicture}
     \caption{Point cloud prediction via sliding (no raycasting) in ZY plane from a hypothetical location using a digital elevation model.}
    \label{Figure:AircraftHypotheticalPoseAbove}
\end{figure}

\begin{figure}[!htb]
    \centering
    \begin{tikzpicture}[x=0.75pt,y=0.75pt,yscale=-0.5,xscale=0.5]
        %uncomment if require: \path (0,542); %set diagram left start at 0, and has height of 542

        %Snip Round Single Corner Rect [id:dp7813370654151424] 
        \draw  [color={rgb, 255:red, 184; green, 233; blue, 134 }  ,draw opacity=1 ][fill={rgb, 255:red, 184; green, 233; blue, 134 }  ,fill opacity=1 ][dash pattern={on 3.75pt off 3pt on 7.5pt off 1.5pt}] (-1,369.5) .. controls (-1,356.52) and (9.52,346) .. (22.5,346) -- (175.5,346) -- (199,369.5) -- (199,463.5) -- (-1,463.5) -- cycle ;
        %Snip Round Single Corner Rect [id:dp9725433180009626] 
        \draw  [color={rgb, 255:red, 184; green, 233; blue, 134 }  ,draw opacity=1 ][fill={rgb, 255:red, 184; green, 233; blue, 134 }  ,fill opacity=1 ] (199,431.5) .. controls (199,427.08) and (202.58,423.5) .. (207,423.5) -- (691,423.5) -- (699,431.5) -- (699,463.5) -- (199,463.5) -- cycle ;
        %Rounded Rect [id:dp8353135865331038] 
        \draw  [fill={rgb, 255:red, 0; green, 0; blue, 0 }  ,fill opacity=1 ] (342.08,217.87) .. controls (342.55,216.93) and (343.7,216.55) .. (344.63,217.03) -- (437.65,263.99) .. controls (438.59,264.47) and (438.96,265.61) .. (438.49,266.54) -- (435.92,271.63) .. controls (435.45,272.57) and (434.3,272.95) .. (433.37,272.47) -- (340.35,225.51) .. controls (339.41,225.03) and (339.04,223.89) .. (339.51,222.96) -- cycle ;
        %Shape: Circle [id:dp7170808088460897] 
        \draw  [fill={rgb, 255:red, 74; green, 74; blue, 74 }  ,fill opacity=1 ] (376.5,244.75) .. controls (376.5,237.85) and (382.1,232.25) .. (389,232.25) .. controls (395.9,232.25) and (401.5,237.85) .. (401.5,244.75) .. controls (401.5,251.65) and (395.9,257.25) .. (389,257.25) .. controls (382.1,257.25) and (376.5,251.65) .. (376.5,244.75) -- cycle ;
        %Straight Lines [id:da5511385616176171] 
        \draw [color={rgb, 255:red, 208; green, 2; blue, 27 }  ,draw opacity=1 ]   (112,378.5) -- (389,244.75) ;
        %Straight Lines [id:da511896702806423] 
        \draw [color={rgb, 255:red, 208; green, 2; blue, 27 }  ,draw opacity=1 ]   (389,244.75) -- (446,456.5) ;
        %Straight Lines [id:da3069520812657579] 
        \draw  [dash pattern={on 0.84pt off 2.51pt}]  (221,162.5) -- (527,313.5) ;
        %Straight Lines [id:da19928498335738898] 
        \draw [color={rgb, 255:red, 0; green, 0; blue, 255 }  ,draw opacity=1 ][line width=1.5]  [dash pattern={on 1.69pt off 2.76pt}]  (221,162.5) -- (143.03,317.02) -- (112,378.5) ;
        %Straight Lines [id:da8105586842351379] 
        \draw [color={rgb, 255:red, 0; green, 0; blue, 255 }  ,draw opacity=1 ][line width=1.5]  [dash pattern={on 1.69pt off 2.76pt}]  (527,313.5) -- (446,456.5) ;
        %Straight Lines [id:da3833966257760052] 
        \draw    (112,378.5) .. controls (110.33,376.83) and (110.33,375.17) .. (112,373.5) .. controls (113.67,371.83) and (113.67,370.17) .. (112,368.5) .. controls (110.33,366.83) and (110.33,365.17) .. (112,363.5) .. controls (113.67,361.83) and (113.67,360.17) .. (112,358.5) .. controls (110.33,356.83) and (110.33,355.17) .. (112,353.5) .. controls (113.67,351.83) and (113.67,350.17) .. (112,348.5) -- (112,347.5) -- (112,347.5) ;
        %Straight Lines [id:da5290793289626303] 
        \draw    (446,456.5) .. controls (444.33,454.83) and (444.33,453.17) .. (446,451.5) .. controls (447.67,449.83) and (447.67,448.17) .. (446,446.5) .. controls (444.33,444.83) and (444.33,443.17) .. (446,441.5) .. controls (447.67,439.83) and (447.67,438.17) .. (446,436.5) .. controls (444.33,434.83) and (444.33,433.17) .. (446,431.5) .. controls (447.67,429.83) and (447.67,428.17) .. (446,426.5) -- (446,424) -- (446,424) ;
        %Straight Lines [id:da14776257776502033] 
        \draw [color={rgb, 255:red, 189; green, 16; blue, 224 }  ,draw opacity=1 ][line width=1.5]  [dash pattern={on 1.69pt off 2.76pt}]  (207,155) -- (112,347.5) ;
        %Straight Lines [id:da5313736165743992] 
        \draw [color={rgb, 255:red, 189; green, 16; blue, 224 }  ,draw opacity=1 ][line width=1.5]  [dash pattern={on 1.69pt off 2.76pt}]  (512,306) -- (446,424) ;
        %Straight Lines [id:da031192920064145313] 
        \draw  [dash pattern={on 0.75pt off 0.75pt}]  (207.71,153.68) .. controls (209.96,152.99) and (211.43,153.78) .. (212.12,156.04) .. controls (212.8,158.29) and (214.27,159.08) .. (216.52,158.4) .. controls (218.77,157.72) and (220.24,158.51) .. (220.93,160.76) -- (221.71,161.18) -- (221.71,161.18)(206.29,156.32) .. controls (208.54,155.64) and (210.01,156.43) .. (210.7,158.68) .. controls (211.39,160.93) and (212.86,161.72) .. (215.11,161.04) .. controls (217.37,160.36) and (218.84,161.15) .. (219.51,163.41) -- (220.29,163.82) -- (220.29,163.82) ;
        
        % Text Node
        \draw (180,259.4) node [anchor=north west][inner sep=0.75pt]  [color={rgb, 255:red, 0; green, 0; blue, 255 }  ,opacity=0.93 ]  {$z_{1}{}_{s}$};
        % Text Node
        \draw (498,375.4) node [anchor=north west][inner sep=0.75pt]  [color={rgb, 255:red, 0; green, 0; blue, 255 }  ,opacity=1 ]  {$z_{2}{}_{s}$};
        % Text Node
        \draw (273,161.4) node [anchor=north west][inner sep=0.75pt]    {$y_{1}{}_{s}$};
        % Text Node
        \draw (472,261.4) node [anchor=north west][inner sep=0.75pt]    {$y_{2}{}_{s}$};
        % Text Node
        \draw (411,208) node [anchor=north west][inner sep=0.75pt]   [align=left] {aircraft hypothetical pose};
        % Text Node
        \draw (134,216.4) node [anchor=north west][inner sep=0.75pt]  [color={rgb, 255:red, 189; green, 16; blue, 224 }  ,opacity=1 ]  {$\hat{z}_{1}{}_{s}$};
        % Text Node
        \draw (445.5,345.4) node [anchor=north west][inner sep=0.75pt]  [color={rgb, 255:red, 189; green, 16; blue, 224 }  ,opacity=1 ]  {$\hat{z}_{2}{}_{s}$};
        % Text Node
        \draw (184,437) node [anchor=north west][inner sep=0.75pt]   [align=left] {DEM};
        \end{tikzpicture}
     \caption{Point cloud prediction via sliding (no raycasting) in ZY plane from a hypothetical location using a digital elevation model.}
    \label{Figure:AircraftHypotheticalPoseBelow}
\end{figure}

Note that the measurements \cref{Eq:pointCloudModel} and the predicted measurements in \cref{Eq:pointCloudPredictionModel,Eq:pointCloudZPredictionModel} are all in body-frame $\mathcal{F}_b$, which coincides with the sensor-frame $\mathcal{F}_s$.

\section{State Estimator}\label{Section:StateEstimator}
The use of an error-state extended Kalman filter \cite[Chapter 3]{grove2013principles} would be appropriate if the measurement model \eqref{Eq:pointCloudPredictionModel} is available analytically in closed-form and can be linearized, which requires gradient info of the terrain profile, as done in SITAN \cite{SITAN_HostetlerAndreas_1983}. The slope info of the terrain profile is less accurate than the terrain profile itself, which puts a limitation on performance. The advent of particle filtering enabled overcoming these measurement function nonlinearities as demonstrated in \cite{SchonMarginalized2005,NordlundMarginalized2009} and \cite{Turan_ParticleFilterTAN}.

Therefore, in this work, instead of the use of an extended Kalman filter, we use particle filtering to estimate the error-state \eqref{eq:INSDynamics_ErrorStateKinematics} in an open-loop integration architecture, because we assume we are unable to inject data into the INS, to ultimately get an estimate of the total-state. The navigation kinematics in \eqref{eq:INSDynamics_ErrorStateKinematics} uses a 15-state dimensional state vector. The lower the dimension of the state vector, the lower is the number of particles we need to use to sample that state-space, which reduces the computational complexity. Therefore, to have efficient particle filters, it is best that the particles sample a lower dimensional state-space, which is the idea behind marginalized particle filters that leverages a linear substructure of the state-space that can be handled using Kalman filters, and marginalizes out the nonlinearities via particle filters.

In our implementation, a marginalized particle filter is used to estimate the error-state, we have the following discrete-time model showing the nonlinear and linear structures using the notation of \cite{SchonMarginalized2005}:

\begin{subequations}\label{eq:MarginalizedModel}
\begin{align}
    s^{n}_{k+1} &= f^n_k(s^{n}_{k}) + A_k^n(s^{n}_{k})s^{l}_{k} + G_k^n(s^{n}_{k}) w^{n}_{k},\\
    s^{l}_{k+1} &= f^l_k(s^{n}_{k}) + A_k^l(s^{n}_{k})s^{l}_{k} + G_k^l(s^{n}_{k}) w^{l}_{k}, \\
    o_{k} &= h_k(s^{n}_{k}) + C_k(s^{n}_{k})s^{l}_{k} + e_{k},
\end{align}
\end{subequations}
where the state of the nonlinear part is $s^{n}_{k} = \delta{p}_k$ and state of the linear substructure is  $s^{l}_{k} = [\delta{v}_k, \delta{\theta}_k, \delta{a}_{\text{bias}_k}, \delta{\omega}_{\text{bias}_k} ]$. Moreover, $o_k$ is the observation or measurement.

A discrete-time representation of \cref{eq:INSDynamics_ErrorStateKinematics} with sampling time $T_s$ can be obtained as shown in \cite[Section 5.4]{sola2017quaternion}, where in our case, we have
\begin{subequations}\label{eq:MarginalizedModelMatrices}
\begin{align}
    f^n_k(s^{n}_{k}) &= s^{n}_{k},\\
    A_k^n(s^{n}_{k}) &=  [T_s I_{3 \times 3}, 0_{3 \times 9}]\\
    G_k^n(s^{n}_{k})  &= I_{3 \times 3},\\
    f^l_k(s^{n}_{k})  &= 0_{12 \times 1},\\
    G_k^l(s^{n}_{k})  &= I_{12 \times 12},\\
    C_k(s^{n}_{k})  &= 0_{(3*M) \times 12}.
\end{align}
\end{subequations}
while $h_k(s^{n}_{k})$ has no analytical closed form representation and represented in \cref{Eq:pointCloudPredictionModel,Eq:pointCloudZPredictionModel}. Moreover,

\begin{equation}
    A_k^l(s^{n}_{k}) = \begin{bmatrix}
        I_{3\times3} & -\bar{R}_{b}^{i}[a_m - \bar{a}_{\text{bias}}]_{\times} T_s & -\bar{R}_{b}^{i} T_s & 0_{3\times3}\\
        0_{3\times3} & \bar{R}_{b}^{i}\{(\omega_m - \bar{\omega}_{\text{bias}})T_s\}^{\intercal}  & 0_{3\times3} & -T_s I_{3\times3}\\
        0_{3\times3} & 0_{3\times3} & I_{3\times3} & 0_{3\times3}\\
        0_{3\times3} & 0_{3\times3} & 0_{3\times3} & I_{3\times3}
    \end{bmatrix},
\end{equation}
where $R\{\omega T_s\}$ indicates the rotation matrix associated with rotation axis of $\omega$ and the associated rotation angle.

\section{Scenarios and Results}\label{Section:Results}
\begin{figure}[!htb]
    \centering
    \begin{subfigure}[b]{0.48\linewidth}
        \centering
        \includegraphics[width=\linewidth]{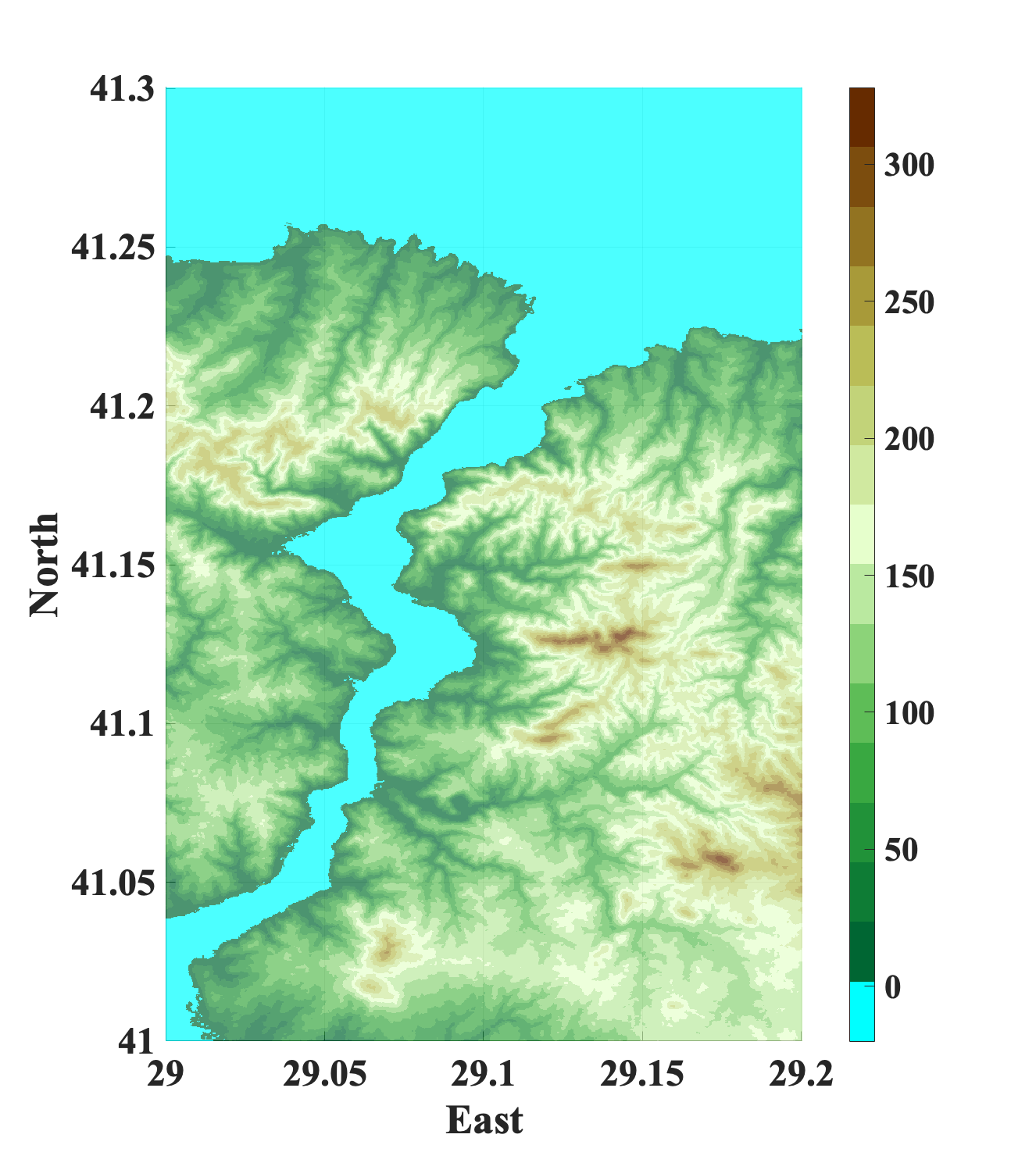}
        \caption{Bosphorus strait terrain.}
        \label{fig:BP_traj}
    \end{subfigure}
    \hfill
    \begin{subfigure}[b]{0.48\linewidth}
        \centering
        \includegraphics[width=\linewidth]{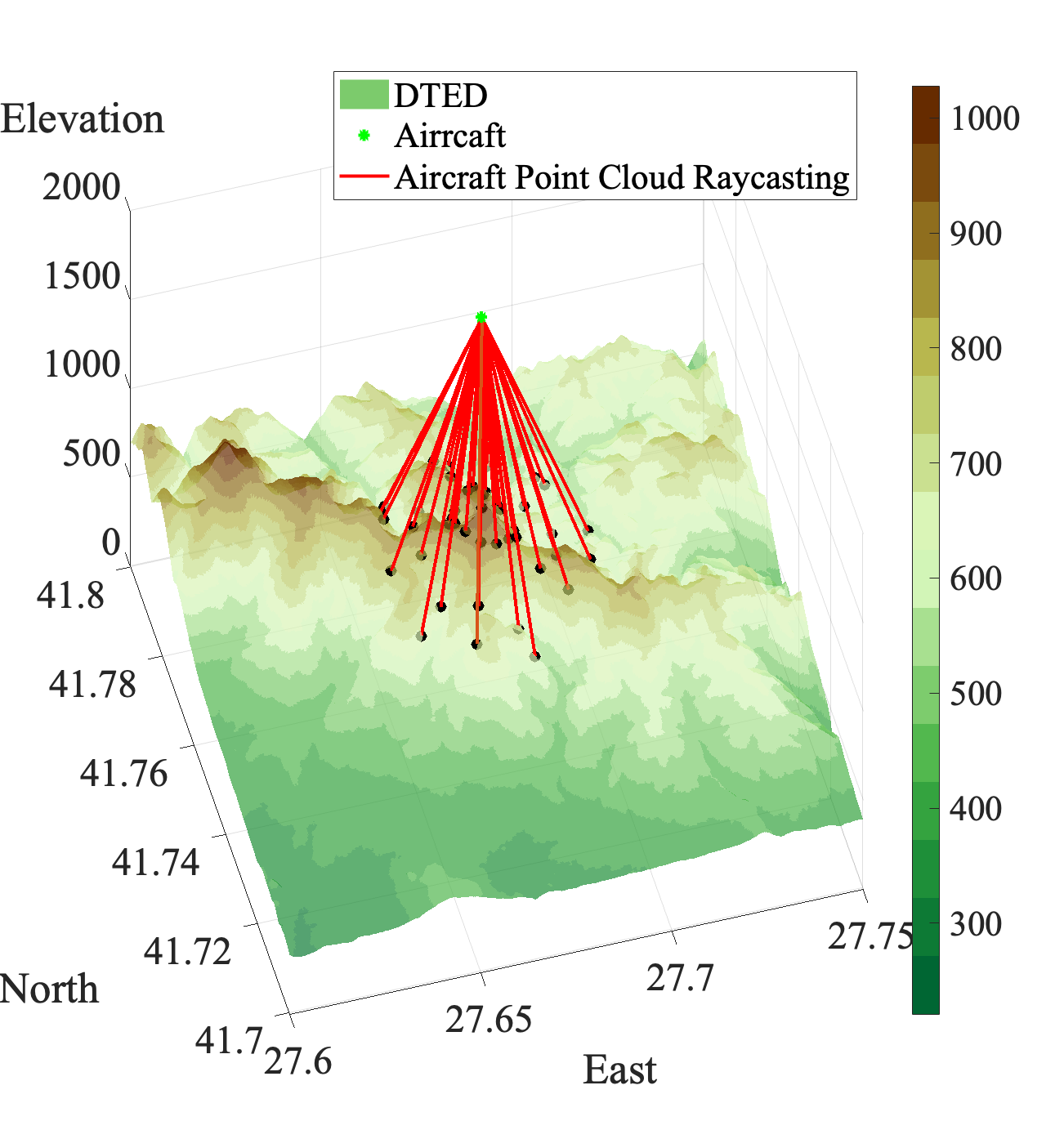}
        \caption{Raycasting over a DEM to generate predicted measurements.}
        \label{fig:PointCloudOverTerrain}
    \end{subfigure}
    \caption{Rugged terrain example and point cloud generation.}
    \label{fig:TerrainAndPointCloud}
\end{figure}

In this section, we apply the state estimator we have discussed along with the terrain measurement models to several scenarios over the rugged terrain of the Bosphorus strait \Cref{fig:BP_traj}. The scenarios involve point cloud measurements comparing raycasting and pattern sliding. We use a DEM that can be obtained from \cite{national_geospatial-intelligence_agency_digital_2022} known also as a Digital Terrain Elevation Data (DTED).

All scenarios share the same common conditions: a high initial estimation error, rugged terrain, 
and a navigation-grade IMU. Two altitude regimes are considered high and low, each tested at a 
different nonzero bank angle, and for each regime both the raycasting and sliding measurement models are 
evaluated, yielding four scenarios in total.

Scenarios S1 and S2 correspond to high altitude with a \SI{30}{\degree} bank angle, using raycasting and 
sliding respectively; their results are shown in \Cref{fig:Raycast_Sliding_highAltitude_lowBank_highE}. 
Scenarios S3 and S4 correspond to low altitude with a \SI{60}{\degree} bank angle, again using raycasting 
and sliding respectively; their results are shown in \Cref{fig:Raycast_Sliding_Low_Altitude_highBank_highE}.

The measurement residual between the \emph{received} and \emph{predicted} point cloud is computed
differently for each model. For raycasting, the residual is computed across all three sensor-frame 
coordinates for each of the $m$ point cloud points according to \cref{eq:RaycastPCerrorSignal}:
    \begin{equation}
    \label{eq:RaycastPCerrorSignal}
    r = \sqrt{
        \Bigl(\tfrac{1}{m}\sum_{i=1}^{m}\!\ \bigl|\,c_{x}(i) - \hat{c}_{x}(i)\bigr|\Bigr)^2
      + \Bigl(\tfrac{1}{m}\sum_{i=1}^{m}\!\ \bigl|\,c_{y}(i) - \hat{c}_{y}(i)\bigr|\Bigr)^2
      + \Bigl(\tfrac{1}{m}\sum_{i=1}^{m}\!\ \bigl|\,c_{z}(i) - \hat{c}_{z}(i)\bigr|\Bigr)^2
    }.
    \end{equation}
For pattern sliding, only the sensor-frame $Z$ coordinate is used, as given by \cref{eq:SlidingPCerrorSignal}:
    \begin{equation}
    \label{eq:SlidingPCerrorSignal}
    r =
    \tfrac{1}{m}\sum_{i=1}^{m}\!\ \bigl|\,c_{z}(i) - \hat{c}_{z}(i)\bigr|.
    \end{equation}

In all scenarios, the IMU update rate and the terrain measurement update rate are assumed to be equal, 
though in general the terrain measurement updates could be run at a lower frequency. All scenarios use 
500 particles.

\begin{figure}[!htb]
    \centering
    \begin{subfigure}[b]{0.48\linewidth}
        \centering
        \includegraphics[width=0.8\linewidth]{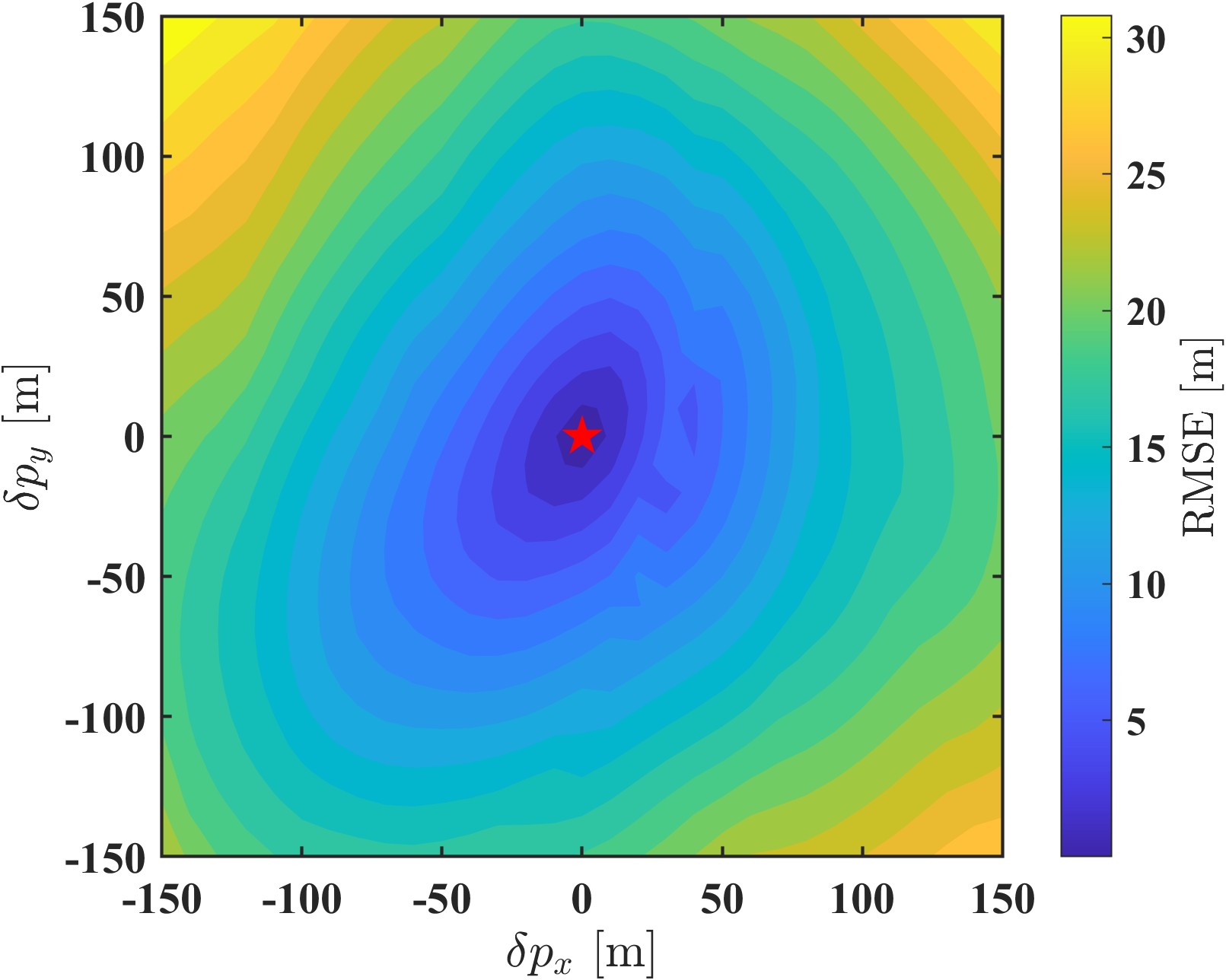}
        \caption{Horizontal position error-state offsets $(\delta p_x, \delta p_y)$.}
        \label{fig:ErrorSurface_xy}
    \end{subfigure}
    \hfill
    \begin{subfigure}[b]{0.48\linewidth}
        \centering
        \includegraphics[width=0.8\linewidth]{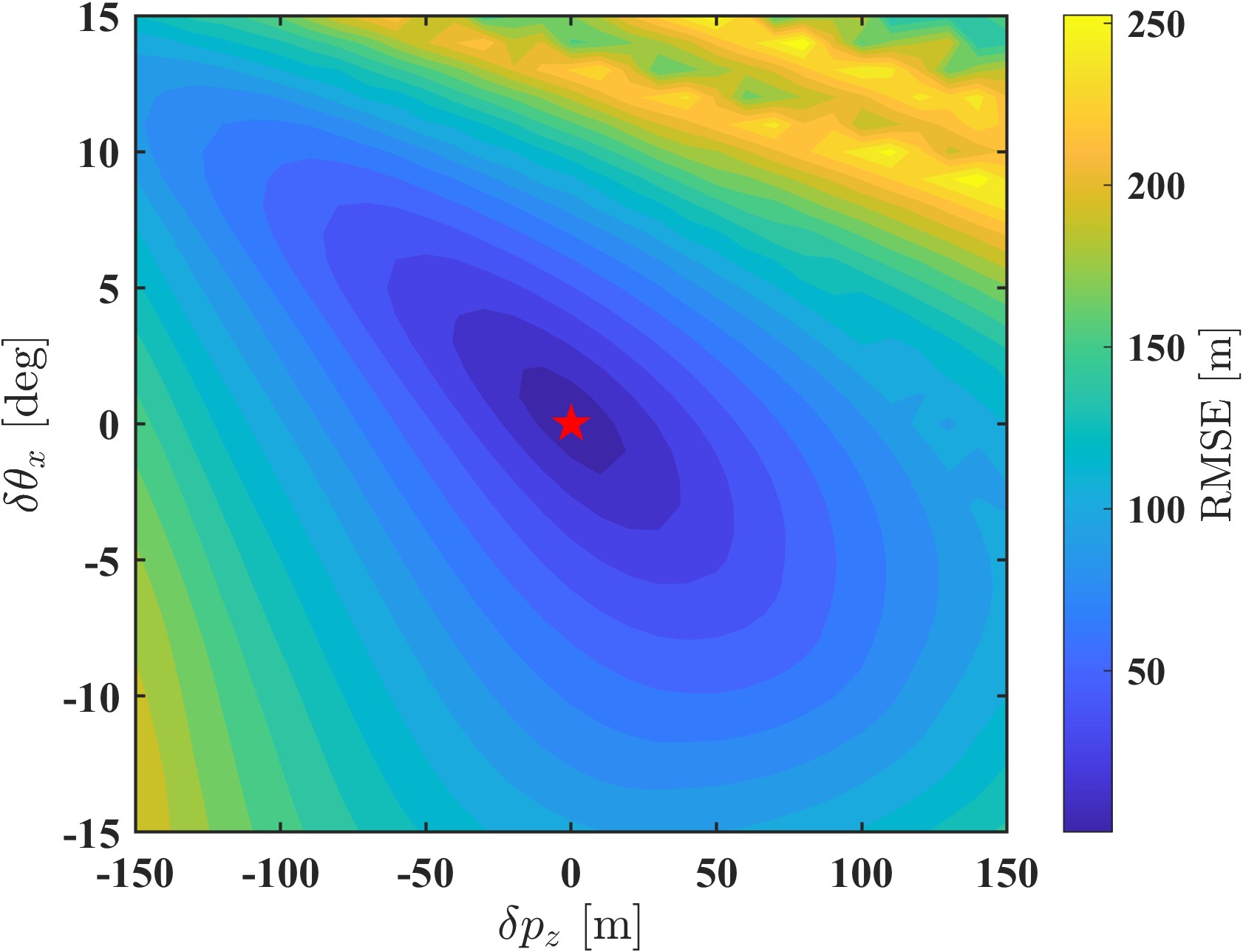}
        \caption{Altitude and bank angle error-state offsets $(\delta p_z, \delta \theta_x)$.}
        \label{fig:ErrorSurface_z_bank}
    \end{subfigure}
    \caption{Point cloud range prediction discrepancy ($\mathrm{RMSE}$) between the sliding grid and raycasting as a function of hypothetical pose error-state offsets around the true aircraft pose (red star).}
    \label{fig:ErrorSurfaceComparison}
\end{figure}

In general, the real-world aircraft is receiving a point cloud measurement which we can emulate via the use of a DEM as shown in \Cref{fig:PointCloudOverTerrain} using a raycasting technique. To quantify the prediction discrepancy between the sliding grid and raycasting across hypothetical particle offsets, we evaluate the root-mean-square range error between the predicted point clouds of both methods over the $M$ scan points:
\begin{equation}\label{eq:SlidingRaycastRangeRMSE}
    e_{\text{range}} = \sqrt{\frac{1}{M}\sum_{i=1}^{M} \Bigl(\bigl\|\hat{c}^{\text{slide}}(i)\bigr\|_2 - \bigl\|\hat{c}(i)\bigr\|_2\Bigr)^2},
\end{equation}
where $\hat{c}(i) \in \mathbb{R}^3$ and $\hat{c}^{\text{slide}}(i) \in \mathbb{R}^3$ denote the $i$-th predicted point in the sensor frame obtained via raycasting and the sliding grid, respectively. \Cref{fig:ErrorSurfaceComparison} illustrates $e_{\text{range}}$ around a true aircraft pose ($z = \SI{1000}{\meter}$, \SI{30}{\degree} bank angle) as a function of horizontal position error-state offsets $(\delta p_x, \delta p_y)$ in \Cref{fig:ErrorSurface_xy} and altitude error-state offset $\delta p_z$ versus roll (bank) attitude error-state offset $\delta \theta_x$ in \Cref{fig:ErrorSurface_z_bank}. At the true aircraft pose ($\delta p = 0$, $\delta \theta = 0$, marked by the red star), the discrepancy between the two methods vanishes, and it increases smoothly as the hypothetical particle pose deviates from the true pose.

\subsection{Raycasting vs. Sliding at High Altitude, Low Banking}\label{Section:Raycast_Sliding_highAltitude_lowBank_highE}

\indent We show results for Scenarios S1--S2 overlapped in \Cref{fig:Raycast_Sliding_highAltitude_lowBank_highE}. As seen in both \Cref{fig:Raycast_Sliding_highAltitude_lowBank_highE} and \Cref{Table:Raycast_Sliding_highAltitude_lowBank_highE}, the use of both ray-casting and pattern sliding achieve similar results. Sliding has the advantage that it is orders of magnitude faster than raycasting as demonstrated in \Cref{fig:RaycastingPatternSlidingComputationalPerformance} discussed later in this work.

The near-identical performance at high altitude can be explained by the relationship between altitude and terrain occlusion. When the aircraft is sufficiently far from the terrain, the scan rays arrive at the surface at near-vertical angles and the terrain features that could occlude one another subtend a smaller angular extent as seen from the sensor. Consequently, the discrepancy between the raycasted surface intersection point and the orthogonal DEM lookup used by the sliding grid diminishes, remaining small across small horizontal and attitude offsets around the true pose as seen in \Cref{fig:ErrorSurface_xy,fig:ErrorSurface_z_bank}. In the limit, as the aircraft altitude increases relative to the terrain relief, the occlusion-induced measurement error of the sliding method approaches zero, and the two measurement models become effectively equivalent. This makes the sliding grid particularly attractive for cruise-altitude operations where its computational savings come at virtually no cost in estimation accuracy.

\subsection{Raycasting vs. Sliding at Low Altitude, High Banking}\label{Section:Raycast_Sliding_Low_Altitude_highBank_highE} 

\begin{figure}[!htb]
    \centering
    \begin{subfigure}[b]{0.48\linewidth}
        \centering
        \includegraphics[width=\linewidth]{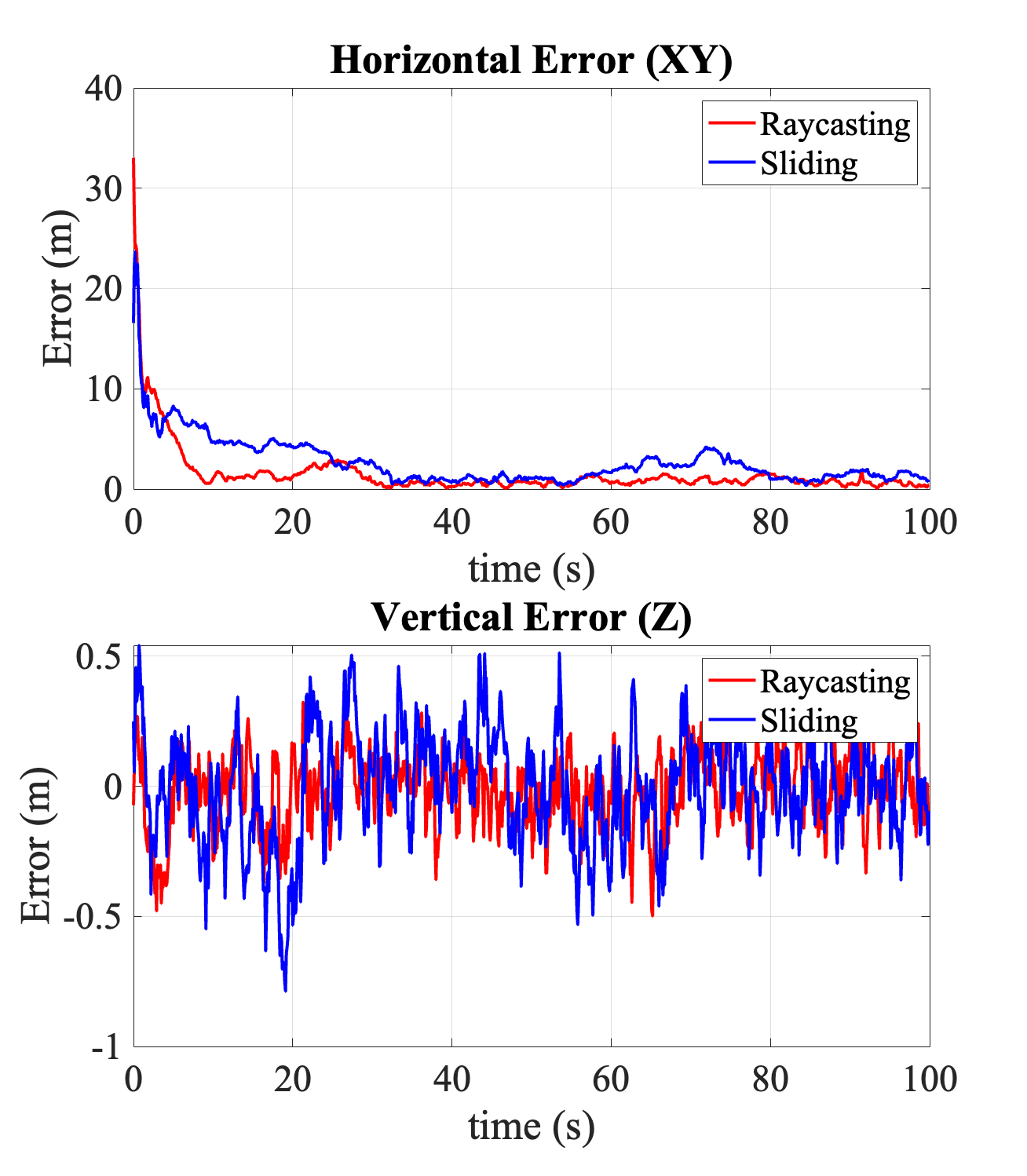}
        \caption{High altitude, \SI{30}{\degree} bank (S1--S2).}
        \label{fig:Raycast_Sliding_highAltitude_lowBank_highE}
    \end{subfigure}
    \hfill
    \begin{subfigure}[b]{0.48\linewidth}
        \centering
        \includegraphics[width=\linewidth]{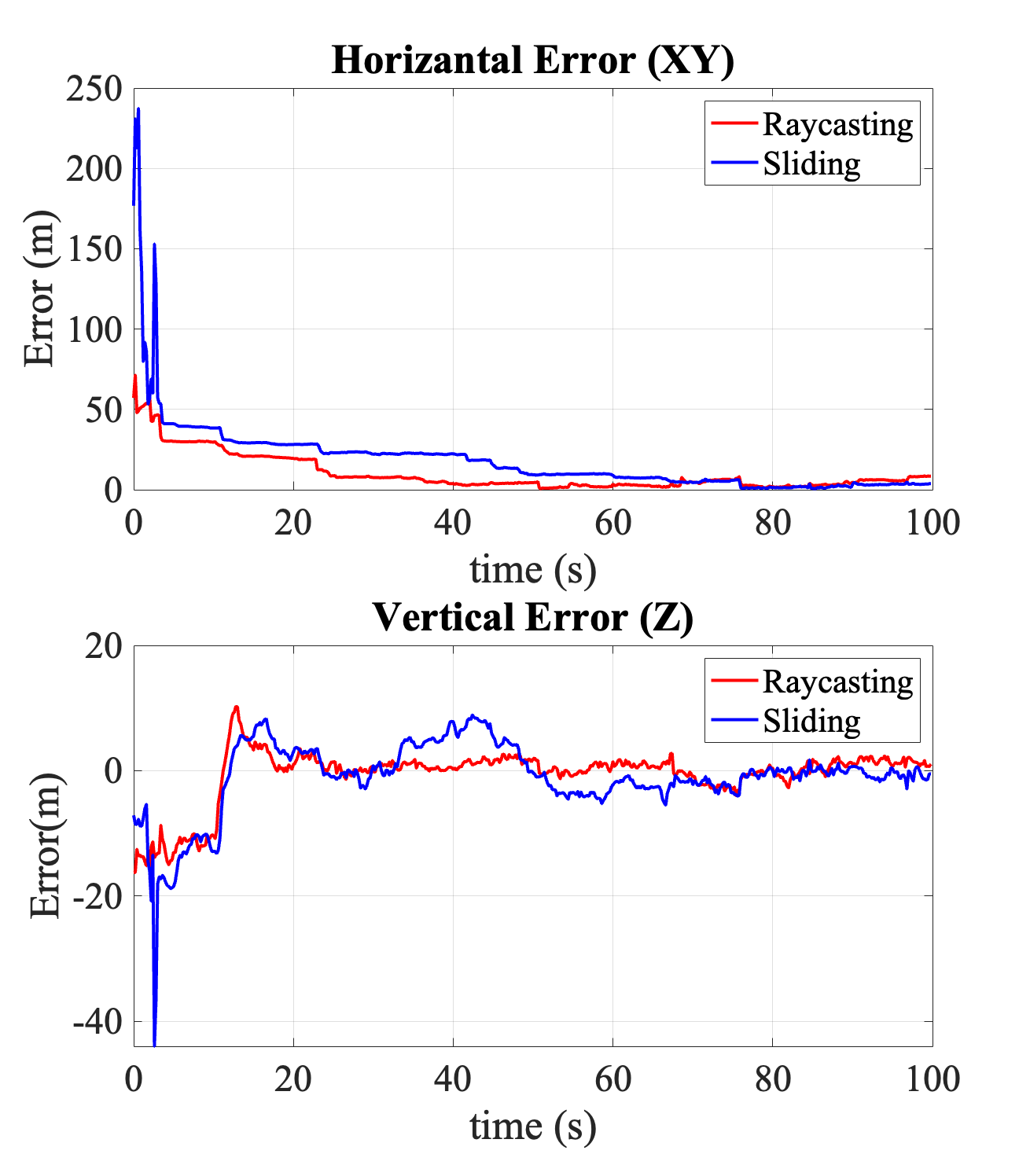}
        \caption{Low altitude, \SI{60}{\degree} bank (S3--S4).}
        \label{fig:Raycast_Sliding_Low_Altitude_highBank_highE}
    \end{subfigure}
    \caption{Comparing the performance of raycasting vs.\ pattern sliding under high initial estimation error, rugged terrain, and navigation grade IMU.}
    \label{fig:Raycast_Sliding_Combined}
\end{figure}

Results for Scenarios S3--S4 are shown in \Cref{fig:Raycast_Sliding_Low_Altitude_highBank_highE} and 
\Cref{Table:Raycast_Sliding_lowAltitude_highBank}. In this scenario, the aircraft is near the top of a hill with a 
\SI{60}{\degree} bank angle, looking toward the non-occluded space opposite the hill. At such a low altitude and high 
bank angle, the sliding grid introduces initial errors due to terrain occlusion and heightened sensitivity to coupled altitude and bank angle offsets (as reflected in \Cref{fig:ErrorSurface_z_bank}) that raycasting naturally handles, leading 
to slower convergence of the state estimator. Nevertheless, the sliding grid does not diverge and eventually converges to comparable performance. This recovery occurs because the aircraft moves away from the hilltop, thereby eliminating the erroneous measurements caused by terrain occlusion. As the geometry becomes more favorable, the filter is able to converge. However, if the aircraft were to remain exposed to this occlusion for a prolonged period, the accumulated measurement errors would likely cause the marginalized particle filter to diverge.

\begin{table}[!htb]
\centering
\caption{RMSE comparison for raycasting vs.\ sliding scenarios.}
\label{Table:Raycast_Sliding_Combined}
\begin{minipage}[t]{0.48\linewidth}
\centering
\subcaption{S1--S2: High altitude, \SI{30}{\degree} bank.}
\label{Table:Raycast_Sliding_highAltitude_lowBank_highE}
\begin{tabular}{llll}
\textbf{System} & \textbf{x(m)} & \textbf{y(m)} & \textbf{z(m)} \\
\hline
\textit{Raycasting} & 2.4483 & 1.9094 & 0.1460 \\
\textit{Sliding}    & 2.4412 & 2.7273 & 0.2116 \\
\end{tabular}
\end{minipage}
\hfill
\begin{minipage}[t]{0.48\linewidth}
\centering
\subcaption{S3--S4: Low altitude, \SI{60}{\degree} bank.}
\label{Table:Raycast_Sliding_lowAltitude_highBank}
\begin{tabular}{llll}
\textbf{System} & \textbf{x(m)} & \textbf{y(m)} & \textbf{z(m)} \\
\hline
\textit{Raycasting} & 8.9555  & 11.4358 & 4.4710 \\
\textit{Sliding}    & 25.1185 & 17.7664 & 5.9862 \\
\end{tabular}
\end{minipage}
\end{table}

\newpage
\subsection{Computational Efficiency}\label{Section:ComputationalEfficiency}
\begin{figure}[!htb]
    \center
    \includegraphics[width=0.5\linewidth]{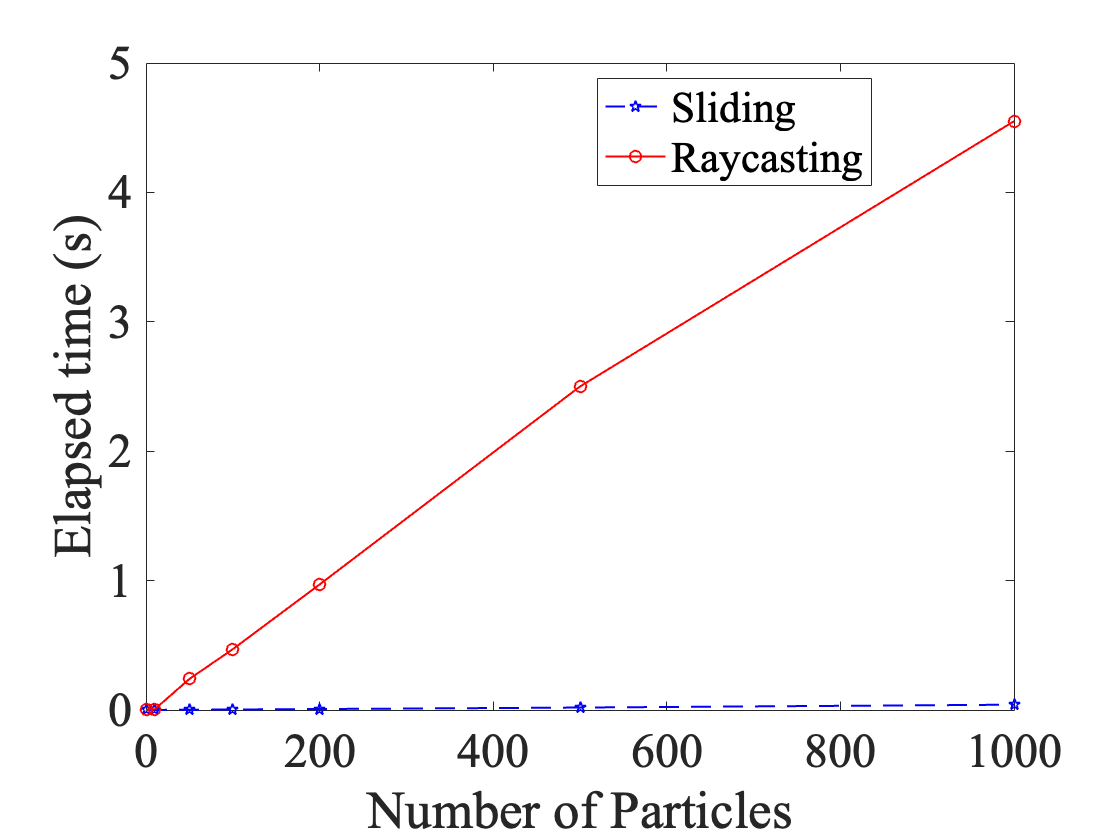}
    \caption{Computational speed comparison: Raycasting vs.\ Pattern Sliding.}
    \label{fig:RaycastingPatternSlidingComputationalPerformance}
\end{figure}

Lastly, the computational efficiency of pattern sliding is significantly better than raycasting as seen in
\Cref{fig:RaycastingPatternSlidingComputationalPerformance}. The plot demonstrates that the sliding grid approach completes its measurement predictions orders of magnitude faster than the raycasting method, scaling much more efficiently as the number of particles increases.

This substantial improvement in execution speed is due to two primary factors:
\begin{itemize}
    \item \textbf{Elimination of ray tracing:} By leveraging the known $x_s-y_s$ pattern of the received point cloud measurement, the algorithm bypasses the need to iteratively cast rays from a source origin and search for surface intersections.
    \item \textbf{Direct orthogonal lookup:} While traversing the predetermined pattern, the expected elevation is obtained directly by querying the DEM at the corresponding horizontal coordinates. This transforms a complex 3D intersection problem into a computationally inexpensive 2D table lookup.
\end{itemize}

As a result, the sliding grid enables the marginalized particle filter to run at high update rates on standard embedded avionics hardware, freeing up critical computational resources for other flight management tasks.
    
\section{Conclusion}\label{Section:Conclusion}
We have demonstrated the use of a novel measurement model, referred to as a sliding grid, for scanning the digital elevation model to predict a point cloud measurement without raycasting. This method leverages the pattern of the received point cloud measurement to avoid computationally expensive raycasting operations, and instead uses a simple reading of the DEM map using horizontal coordinates calculated from the known pattern. The sliding grid achieves comparable localization accuracy to raycasting, especially at higher altitudes, while being orders of magnitude faster computationally. Crucially, by removing the need for raycasting, the sliding grid eliminates the dependency on dedicated GPU hardware or high-performance CPUs that raycasting typically demands, thereby enabling terrain-aided navigation on resource-constrained embedded platforms. At low altitudes near terrain with high bank angles, raycasting provides better initial convergence due to its ability to handle occlusion, however the sliding grid still converges without diverging.

This study serves as a basis for further theoretical analysis and potential flight tests to further validate and elevate the technology readiness level of the methods presented.

\section*{Acknowledgments}
This work was supported by Turkish Aerospace Industries.

\bibliography{refs}

\end{document}